\documentstyle[12pt]{article}

\begin{document}

\begin{titlepage}
\begin{center}
{\Large
\bf
On the Non-invasive Measurement of the Intrinsic Quantum Hall Effect
\vskip 0.5cm
\large
\rm
D.\ P.\ Chu and P.\ N.\ Butcher\\
Department of Physics, University of Warwick,\\
Coventry CV4 7AL, U.K.
}
\end{center}

\vspace{0.2cm}

\begin{abstract}
With a model calculation, we demonstrate that a non-invasive
measurement of intrinsic quantum Hall effect defined by
the local chemical potential in a ballistic quantum
wire can be achieved with the aid of a pair of voltage
leads which are separated by potential barriers from
the wire. B\"uttiker's formula is used to determine the
chemical potential being measured and is shown to reduce
exactly to the local chemical potential in the limit of
strong potential confinement in the voltage leads. Conditions
for quantisation of Hall resistance and measuring local
chemical potential are given.
\end{abstract}

\vspace{0.2cm}

\end{titlepage}

\newpage

To study the electronic transport properties of a system,
it is normal to use at least four leads, attaching
two pairs of leads to the system to measure the current passing
through and the voltage drop across it. In the macroscopic
regime, the scale of system is much larger than the
scale of the measurement leads. Consequently, this approach
has very little effect on the system being measured and the
measurement results can be used to fully characterise the system
itself. This desirable situation has changed with the rapid
development of semiconductor fabrication techniques which
make it possible to investigate two-dimensional electron
gas (2DEG) microstructures \cite{book}. In this case
both current and voltage leads become an inseparable part
of the system being measured. Moreover, the dimensions of
the part being measured and the measurement leads are
of the same order and can be comparable with the de~Broglie
wavelength of the electron propagating in the system. Many novel
phenomena are observed in this situation. They are attributed to
this new partnership between the system being measured and the
leads and are explained successfully by using Landauer-B\"uttiker
formulae which reveal the relationship of resistance to transmission
coefficient between leads \cite{landauer70,buttiker86}.
B\"uttiker has proposed a general formula to determine the chemical
potential measured by a voltage lead through a current-stop
procedure \cite{buttiker86,buttiker88}:
\begin{equation}
\mu_l =
\displaystyle{{T_{ls}\mu_s + T_{ld}\mu_d}\over {T_{ls} + T_{ld}}}
\label{u}
\end{equation}
where $T_{ls}$($T_{ld}$) is the sum of all the transmission
coefficients for a carrier
incident in lead $s$($d$) to be transmitted to lead $l$ and
the subscripts $l$, $s$, and $d$ denote voltage, source, and
drain leads respectively.

A problem comes when one asks how to determine the {\it intrinsic}
resistance of such a microstructure system, {\it i.e.} its {\it own}
response to the change of environment \cite{chu931}. To do this, it
is necessary to study the effect of the leads on the resistance
measurement in detail. The two kinds of leads (current and voltage)
have different interactions with the system. Current leads function
as sources and drains which respectively inject electrons into and
collect them from the system being measured.
Voltage leads do not have any net electron exchange with
the system; they determine the potential being measured by a
current-stop procedure. Furthermore, different shapes of leads
give different results. To be definite we consider ideal current
leads, {\it i.e.} hard wall ballistic electron
waveguides which become an integral part of a system. (They are
used as filters \cite{maslov93} to get rid of fluctuations,
evanescent modes, etc. coming from the reservoirs
and introduce standard propagating modes of
electrons into the system.) The injection modes are determined
by the character of the ideal leads only. If the shape
of the current leads are fixed we just need to concern about the
effect of the voltage leads.

There is no confusion as long as we use the one pair of leads to
measure the current passing through and the ``voltage drop''
across a system. The result of such measurements is a conventional
longitudinal resistance. When we have separate pairs of leads to
measure current and ``voltage drop'', which is essential in
studies of the quantum Hall effect (QHE), the resistance measured
reflects the behaviour of the original system plus a pair of voltage
leads and is voltage-lead dependent. We are not able to isolate the
contribution from the system being measured in the total signal.
The only way to solve this problem is to reduce the coupling between
the system to be measured and the voltage leads. The weaker the
coupling, the less the measurement result is effected by the
measurement process. However, we know that there is no way to measure
a system without {\it some} perturbation of the system being measured.
What we must do is to make the coupling small enough so that the
measured resistance does not change within the accuracy of the
measurement instrument when the coupling decreases further. Then,
in this sense, the measurement is non-invasive and the measured
resistance can be regarded as intrinsic to the system which we
measure.

Many papers discuss methods of making a non-invasive voltage leads.
Most of the works are carried out at the geometrical edge of the 2DEG
microstructure due to technical reason. Li and Thouless suggested
using a scanning-tunnelling-microscope tip as a weakly coupled voltage
lead to detect the electrostatic potential response of QHE from an
etched edge \cite{li90}. Field {\it et al.} use a
separate quantum point contact sited at the side of a gated edge
to achieve non-invasive measurement of electrostatic potential
\cite{field93}. When we work out the resistance of a system, however,
we need to know the chemical potential difference rather the
electrostatic potential difference between two points, as is stressed
by Engquist and Anderson \cite{engquist81}. Experimental attempts to
measure resistance in the weak coupling limit have been made recently
by Shepard {\it et al.} \cite{shepard92}. It is much more difficult to
determine the chemical potential at a certain point of a transport system.
The main reason is that the chemical potential in a system is normally
not well defined when there is a net current flowing through it. Many
suggestions have been made about how to define this quantity locally in
a system away from thermal equilibrium
\cite{engquist81,entin86,butcher93}.
They all lead to the same chemical potential
and average electron occupation in an equilibrium system
as has been pointed out by Landauer \cite{landauer75,landauer88}.
Different procedures for non-invasive measurement have been suggested,
{\it e.g.} phase-insensitive \cite{engquist81} and phase-sensitive
\cite{buttiker90}. They give different results when there is a net
current passing through the system with reflections.

To avoid of these problems, a particular formula has been introduced
through an assumption of a virtual contact measurement procedure for
both single and multi-mode two-terminal cases by Entin-Wohlman
{\it et al.} \cite{entin86} and Imry \cite{imry89} respectively.
The advantage of this formula is that it defines a local chemical
potential (LCP) in a non-equilibrium system so that we can {\it calculate}
the resistance between any two points in a system in which
net currents are flowing without introducing voltage leads.
B\"uttiker derives a similar expression for a self-consistent
electrostatic potential \cite{buttiker88}. The same formula for the
LCP is obtained in a general multi-mode and multi-terminal case by
using only the assumptions inherent in Landauer-B\"uttiker formulas
\cite{butcher93}. It is
\begin{equation}
\mu ({\bf r}) ={{\displaystyle\sum_t p_t \mu_t} \over
{\displaystyle\sum_t p_t}}
\label{c}
\end{equation}
where $ p_t = \sum_m \left| \psi_{tm}({\bf r}) \right|^2 / v_{tm} $.
Here $t$ labels the leads feeding the microstructure, $v_{tm}$ is
the group velocity of mode $m$ in lead $t$ and
$\psi_{tm}({\bf r})$ is the total wave function generated by
an incident wave of unit amplitude in mode $m$ in lead $t$.
We would like to stress that the LCP is phase-sensitive. The phase
relation between the incident wave and the reflected wave is
fully considered in the calculation of the wave function for the
whole system. Moreover, the resistance determined by the LCP is
non-local resistance which is not normally additive.

In this paper, we model a non-invasive measurement procedure in
a system consisting of a quasi-one-dimensional ballistic
quantum wire (BQW) and two voltage leads.
The current leads are part of the BQW.
Transmission coefficients are calculated and the chemical potential
as well as the Hall resistance associated with it
are obtained using B\"uttiker's formula, Eq.~(\ref{u}). In the
strong confinement limit, we prove analytically that B\"uttiker's
formula is equivalent to the formula for the LCP, Eq.~(\ref{c}),
and the Hall resistance approaches the intrinsic
Hall resistance defined by the LCP \cite{chu931}.
Numerical results are given to show how the character of the
voltage leads affects the Hall resistance and to which every mode
therein makes a non-negligible contribution.

The main part of our model system is a non-interacting 2DEG with
electron density $n_s$ which is confined in a space of width
$W$ in the $x$-$y$ plane by infinite potential barriers at
$y=\pm W/2$. The two ends of the BQW are connected to
the electron reservoirs with chemical potential $\mu_s$
(at the end where $x<0$) and $\mu_d$ (at the end where $x>0$)
respectively. When $\mu_s \neq \mu_d$, there is a net
current traversing the BQW. To model a four-terminal measurement
of the Hall resistance, we use the weak-link model studied
by Peeters \cite{peeters88} and later by Akera and Ando
\cite{akera89} to put two voltage leads on the two sides
of the wire in the $x$-$y$ plane and parallel to the $y$-axis.
The confinement potential in the voltage leads has the form of
$m^*\omega_p^2x^2/2$ and is characterised by an equivalent
magnetic field $B_p=m^*\omega_p/e$ where $m^*$ is the
effective mass of electron. We assume these two types of
confinements for the BQW and the voltage leads respectively
because they are mathematically simple and are close to the
calculated self-consistent potential profiles for the relatively
wide and narrow BQWs in which the Fermi energy is the same
\cite{laux88} which is the case when we explore the strong
confinement limit in the voltage leads. Moreover, two identical
tunnelling barriers with heights $V_b$ and widths $b$ are
symmetrically placed between the wire and the ends of the
voltage leads. The amount of current leaking into the voltage
leads can be made very small by increasing the product $V_bb$
so that we can approach the non-invasive limit defined above.
For convenience in the calculations, delta functions of area
$V_b b$ are located at $y=\pm W/2$ to describe potential
energies of the tunnelling barriers as $V_b b \delta(y\mp W/2)$.

A magnetic field $B$ is applied in the direction perpendicular
to the $x$-$y$ plane and is described in the Landau gauge by writing
the vector potential as ${\bf A}=(-By,0,0)$ for the BQW and
as ${\bf A}=(0,Bx,0)$ for the voltage leads. Taking account of
the gauge difference between the two regions,
the tunnelling wave function of an electron from the BQW to
the voltage lead at $y=W/2+\epsilon$ ($\epsilon \rightarrow 0^+$)
is
\begin{equation}
\psi^{(n\pm )}(x,\displaystyle{W\over 2}) = C^{(n\pm )}_{\psi}
\exp \left[ \displaystyle{i \left( \pm k^{(n)}_x
            +{W\over {2l^2_c}} \right) x} \right]
\label{psi}
\end{equation}
with
$ C^{(n\pm )}_{\psi}=
-\displaystyle{{\hbar^2 \over {2m^*}} {1 \over {V_bb}}}
\left.
\displaystyle{
{d\chi^{(n)} \left( \displaystyle{y
            \over {l_c}}\mp l_ck^{(n)}_x \right) }
\over {dy}}
\right|_{y=W/2}, $
$l^2_c=\hbar/e|B|$, and $\chi^{(n)}$ for the $n$-th eigenfunction
for an electron in the BQW. The Fermi wave vector $k^{(n)}_x$ is
all real and positive and determined with the Fermi energy
$E_F$ by a sum constrained to keep $n_s$ fixed \cite{chu931}.
The $\pm$ sign refers the mode propagating along $\pm x$
direction.

The eigenfunction of electron in the voltage lead at $y> W/2$ is
\begin{equation}
\phi^{(m)}(x,y)= C^{(m)}_{\phi}
                 e^{-\displaystyle{1\over 2}{\eta_m}^2}
                 H_m(\eta_m)
\end{equation}
with
$
C^{(m)}_{\phi}=\displaystyle{{(1+\gamma^2)^{1/8}}
\over {(2^m m! \pi^{1/2} l_c)^{1/2}}}
     \exp \left[
     {\displaystyle{ik_y^{(m)} \left(y-{W \over 2}
     \right)}} \right] ,$
$\eta_m=(1+\gamma^2)^{1/4}\displaystyle{x \over {l_c}}
          +\displaystyle{ 1\over {(1+\gamma^2)^{3/4}} }
                        l_ck^{(m)}_y ,$
where $m$ is the mode index, $H_m$ is Hermite polynomial, $k^{(m)}_y$
is the Fermi wave vector of the electron, and $\gamma=B_p/B$. The
Fermi wave vector $k^{(m)}_y$ is either real or imaginary
(corresponding to propagating and evanescent mode) due to the
parabolic potential confinement.

We choose the $k^{(m)}_y$'s so that the electron energy
in the voltage lead is $E_F$ when we expand the tunnelling electron
wave function in Eq.~(\ref{psi}) in the terms of
the eigenfunction of electron in the voltage lead at $y=W/2$
\begin{equation}
\psi^{(n\pm )}(x,\displaystyle{W\over 2})=
\displaystyle\sum_m g^{(n\pm )}_m
\phi^{(m)}(x,\displaystyle{W\over 2})
\end{equation}
The wave functions $\phi^{(m)}(x, W/2)$ are
normalised but they are not orthogonal. Consequently,
the $g^{(n\pm )}_m$ are determined by following equations:
\begin{equation}
\displaystyle\sum_m f_{jm} g^{(n\pm )}_m = h^{(n\pm )}_j
\label{g}
\end{equation}
with $f_{jm}=\int^{+\infty}_{-\infty} dx
\phi^{(j)}(x, W/2) \phi^{(m)}(x, W/2) ,$
$h^{(n\pm )}_j= \int^{+\infty}_{-\infty} dx
\phi^{(j)}(x, W/2) \psi^{(n\pm)}(x, W/2).$

After solving Eq.~(\ref{g}), we can directly calculate
the transmission coefficients from their definitions
\begin{equation}
\begin{array}{c}
T_{ls} = \displaystyle\sum_n T^{(n)}_{ls}
       = \displaystyle\sum_n \displaystyle{1\over {v_n}}
         \displaystyle\sum_m v_m \left| g^{(n+)}_m \right|^2 , \\
T_{ld} = \displaystyle\sum_n T^{(n)}_{ld}
       = \displaystyle\sum_n \displaystyle{1\over {v_n}}
         \displaystyle\sum_m v_m \left| g^{(n-)}_m \right|^2 , \\
\end{array}
\label{tr}
\end{equation}
where the summations over $n$ and $m$ include all the values for
which $\{n|E_F=E(k^{(n)}_x)\}$ and $\{m|k^{(m)}_y\in {\rm R}\}$
respectively. Here, the subscripts have the same meaning as in
Eq.~(\ref{u}) and $v_m\geq 0$ ($v_n\geq 0$) is the group velocity of
electron of the $m$-th ($n$-th) propagating mode in the voltage
lead (BQW) with its energy equals $E_F$.

The chemical potential defined by B\"uttiker's formula, Eq.~(\ref{u}),
can be calculated easily from $T_{ls}$ and $T_{ld}$ in Eq.~(\ref{tr}).
If we have strong potential confinement in the voltage
lead, {\it i.e.} $B_p\gg B$, the leading term of the coefficients
$f_{jm}$ and $h^{(n\pm)}_j$ are
$$
f_{jm}\simeq \delta_{jm}, \ \
h^{(n\pm)}_j\simeq
\displaystyle{{(2\pi )^{1/2}l_c}\over {\gamma^{1/2}}}
  i^j C^{(j)}_{\phi} C^{(n\pm)}_{\psi} H_j(0).
$$
Consequently, we can easily show that
$$
\begin{array}{rcl}
\displaystyle\sum_m v_m \left| g^{(n\pm)}_m \right|^2 &\simeq
&\displaystyle\sum_m v_m \left| h^{(n\pm)}_m \right|^2 \\
{}&= & {\rm const.}\times \gamma^{-1/2} \left| C^{(n\pm)}_{\psi}
\right|^2. \\
\end{array}
$$
Hence, the chemical potential measured by the voltage lead attached
to the BQW at the edge $y=W/2$, which is defined by Eq.~(\ref{u}),
reduces to
\begin{equation}
\begin{array}{rcl}
\mu_l &=&
\displaystyle{
  {\displaystyle\sum_n \displaystyle{1\over v_n} \left(
\left| F^{(n+ )} \right|^2 \mu_s
+
\left| F^{(n- )} \right|^2 \mu_d
\right) }
\over
  {\displaystyle\sum_n \displaystyle{1\over v_n} \left(
\left| F^{(n+ )} \right|^2
 +
\left| F^{(n- )} \right|^2
 \right)}
 } \\
\end{array}
\label{p}
\end{equation}
where
$F^{(n\pm )} = \left. \partial\psi^{(n\pm )}(x,y) / \partial y
\right|_{y=W/2}$.

We see by inspection of Eq.~(\ref{p}) that B\"uttiker's chemical
potential is identical to the LCP at $y=W/2$ defined by Eq.~(\ref{c})
in a BQW with no voltage probes attached. It is important to note
that the eigenfunctions of electrons in the voltage lead do not
change significantly as we change $B$ when $B_p\gg B$. In this
situation, the coupling strength of each mode in the BQW
to the voltage lead will depend only on the character of
the mode itself and nothing else. As long as these electron
modes are undisturbed by the voltage lead, we can make a
non-invasive measurement and the LCP defined by Eq.~(\ref{c})
and the intrinsic Hall resistance associated with it are
the quantities being measured.

The Hall resistance $R_H$ associated with the chemical potential
$\mu_l$ defined by B\"uttiker's formula Eq.~(\ref{u})
is obtained by solving Eq.~(\ref{g}) with $m^*=0.068\ m_e$ for GaAs,
$W=100\ {\rm nm}$, and $n_s=1.1\times 10^{15}\ {\rm m}^{-2}$ so that
three subbands are populated when $B=0$. We include the
necessary number of evanescent modes in the voltage leads, such
that no change of the expansion coefficient $g^{(n\pm )}_m$
(for the $k_y^{(m)}\in {\rm R}$) occurs when we take more evanescent
modes into account. The same zero point of potential is used
in both the BQW and the voltage leads.

Fig.~1 shows the changes of the dependence of $R_H$ on $B$ from
$B_p\sim B$ to $B_p\gg B$. The solid line is the result of $R_H$
associated with the LCP of Eq.~(\ref{c}) as studied in
Refs.~\cite{chu931,chu932}, while the dashed line
and the dot-and-dash line are the results of $R_H$ calculated
by B\"uttiker's formula Eq.~(\ref{u}) for $B_p=1\ {\rm T}$ and
$11\ {\rm T}$ respectively. The dotted line is for the longitudinal
resistance which is perfectly quantised since there are no
reflections in the BQW. We verify Form Fig.~1 that the $R_H$
derived from B\"uttiker's formula does approach the intrinsic
Hall resistance derived from the LCP when we increase $B_p$ and has it
as its limit when $B_p\gg B$. In the range of $0 < B < 0.6\ {\rm T}$,
there is a quenching of $R_H$ for both $B_p=1\ {\rm T}$
and $B_p=11\ {\rm T}$ as we have found for the intrinsic quantum
Hall resistance in the BQW with interacting electrons \cite{chu932}
and the magnitude of $R_H$ reduces when $B_p$ increases. The dips
of $R_H$ are deeper than that displayed in
Ref.~\cite{chu932} because only one electron state is used here
to calculate $R_H$ rather than the electron states in a small
but finite range of energy due to the chemical potential difference
between source and drain. We also notice from Fig.~1 is that there
is a quantised plateau on the $R_H$ curve around $B\sim 2.2\ {\rm T}$
when $B_p=1\ {\rm T}$ instead of the dip found when $B_p=11\ {\rm T}$.
This implies that measurements of $R_H$ made with two weakly
confined voltage leads give results which are similar to those
found using a Hall bar geometry. On the other hand, we have confirmed
both analytically and numerically, that strongly confined voltage
leads give the $R_H$ values predicted by the LCP given in
Eq.~(\ref{c}).

In Fig.~2, we present results for the single mode form factor
$F^{(n)}=(T^{(n)}_{ls}-T^{(n)}_{ld})/(T^{(n)}_{ls}+T^{(n)}_{ld})$
as defined in Ref.~\cite{peeters88} for propagating modes.
Figs.~2(a) and 2(b) are for $B_p=1\ {\rm T}$ and $11\ {\rm T}$
respectively. The dotted, dashed, dot-and-dash, and solid lines
are for $F^{(1)}$, $F^{(2)}$, $F^{(3)}$, and the total form factor
$F=\sum_n (T^{(n)}_{ls}-T^{(n)}_{ld}) /
\sum_n (T^{(n)}_{ls}+T^{(n)}_{ld})$ respectively. Each mode (not
only the one closest to the edge of the BQW) makes a contribution
to the total form factor. Quantisation of $R_H$ can be reached
when every single mode form factor $F^{(n)}=1$. Comparing
Figs.~2(a) and 2(b), we can see that $F^{(n)}$ is closer to $1$
when $B_p\sim B$ than when $B_p\gg B$. In other words, better
quantisation plateaus of $R_H$ can be observed by using more
loosely confined voltage leads.

In summary, we investigate the possibility of making non-invasive
measurement of the LCP and the intrinsic quantum Hall resistance.
A model procedure is used for calculation. We proved that the chemical
potential described by B\"uttiker's formula Eq.~(\ref{u}) has the LCP
defined by Eq.~(\ref{c}) as its limit when the potential confinement
parameter $B_p$ in the voltage leads increases indefinitely.
Numerical calculations are carried out, which confirm the limiting
behaviour of the quantum Hall resistance $R_H$. Quenching of
$R_H$ is seen in a broad range of $B_p$. Our calculations indicate
that it is possible to measure the LCP given in Eq.~(\ref{c}) and the
intrinsic QHE non-invasively in some circumstances. It is hoped that
further experimental studies of this challenging problem will be made.

This work was supported by the United Kingdom Science and
Engineering Research Council.

\newpage
\noindent{\Large \bf Figure Caption}

\noindent Fig.\ 1 \/\/\/ Hall resistance $R_H$ calculated from
Eq.~(\ref{u}) when $B_p=1\ {\rm T}$ (dashed line) and $11\ {\rm T}$
(dot-and-dash line). The $R_H$ associated with LCP and the
longitudinal resistance of BQW are shown by solid line and dotted
line respectively.

\vskip 0.5cm
\noindent Fig.\ 2 \/\/\/ The total form factor $F$ (solid line) with
three single mode form factor $F^{(1)}$ (dotted line), $F^{(2)}$
(dashed line), and $F^{(3)}$ (dot-and-dash line) for (a) $B_p=1\ {\rm T}$
and (b) $B_p=11\ {\rm T}$.

\end{document}